# Astronomical Infrared Bands and Diffuse Interstellar Bands Both Reproduced by Hydrocarbon Pentagon-Hexagon Combined PAH Molecules

Norio Ota

Graduate school of Pure and Applied Sciences, University of Tsukuba, *1-1-1 Tennodai, Tsukuba-City Ibaraki 305-8571, Japan*

This study theoretically predicts the specific Polycyclic Aromatic Hydrocarbon (PAH) molecules to reproduce both astronomically observed Infrared Bands (IR) and Diffuse Interstellar Bands (DIB). In our recent paper, we could reproduce IR by the hydrocarbon pentagon-hexagon combined PAH molecules using Density Functional Theory (DFT). Found molecules were ($C_{53}H_{18}$), and ($C_{23}H_{12}$) with two carbon pentagons among hexagon networks. Origin of DIB may come from the molecular orbital excitation. We applied Time-Dependent DFT calculation. In case of ($C_{53}H_{18}$), by comparing calculation with observed DIB, we found 7 coincide bands among 42 calculated bands within observed band width. For example, neutral ($C_{53}H_{18}$) shows calculated 577.35nm band coincide well with observed DIB at 577.95nm within 1.55nm observed band width. Mono-cation shows calculated 627.87nm correspond to observed 627.83nm, also for di-cation calculated 635.89nm to observed 635.95nm. For smaller size molecule ($C_{23}H_{12}$), we found 5 coincide bands, of which mono-cation shows calculated 713.92nm coincide with observed 713.80nm, di-cation shows calculated 653.27nm correlate to observed 653.21nm. By such quantum-chemical survey, we could predict specific PAH molecules floating in interstellar space.

## 1. Introduction

This paper theoretically predicts background molecules to coincide with both astronomically observed infrared emission bands (IR) and Diffuse Interstellar absorbed Bands (DIB)[1]. In our recent paper[2], specific PAHs were reported to coincide well with observed IR, which were selected under the astronomical top-down material creation model[3] by the density functional theory (DFT) based calculation on polycyclic aromatic hydrocarbon (PAH) molecules. Obtained PAHs were ($C_{53}H_{18}$), ($C_{23}H_{12}$) and ($C_{12}H_8$), which has one or two pentagon hydrocarbon units among hexagon networks. In this study, we expected that those molecules may contribute again on DIB. It was supposed that DIB may come from the photoexcitation (absorption) between molecular orbitals. We tried the Time-Dependent DFT (TD-DFT) calculation.

To identify specified PAH, we did paradigm shift from the bottom-up astronomical dust creation scheme to the top-down one. Those were reviewed by A. Tielens[3] in 2013. Conventional bottom-up process was that few atoms create larger molecule one by one as like laboratory chemical reaction. For example, $C_2$ chemically reacts with $C_4$ to result $C_6$. On the other hand, the top-down creation was physical transformation of larger molecule to smaller one as like laboratory physical sputtering, for example from $C_6$ to $C_5$ by high-speed proton attack on carbon atom. We already applied this top-down scheme to molecular vibrational IR and show successful coincidence[2] with well-known astronomical PAH bands in a wide wavelength from 3.2, 6.3, 7.7, 8.6, 11.2, to 12.7 micrometer[4)-11)]. Recent review on astronomical PAHs was opened by A. Li [12] in 2020.

The Diffuse Interstellar Bands (DIB) are a large set of absorption features at a wide range of visible to infrared wavelengths to associate partly with carbon and /or hydrocarbon molecules as reviewed by T. R. Geballe[1]. The discovery of the first diffuse interstellar bands was made by Mary Lea Heger[13], almost 100 years ago. Until now, there observed over 500 DIBs[14)-19)]. However, any specific molecule had not been identified yet. In 2015, four near-infrared DIBs were matched to the laboratory spectrum of singly cationic fullerene C60+ [20)-23)]. However, others were not identified yet. This is the great mystery for the astrochemistry and molecular science.

Purpose of this study is to indicate the specific molecule to satisfy both astronomically observed IR and DIB.

## 2. Top-Down Scheme for Creating Interstellar Molecule

We followed astrophysical dust creation model. The first assumption is the creation of giant molecule after the death of large star. Just after star explosion, high temperature carbon plasma was released to circumstellar space as imaged in (1) of Fig. 1. There causes cooling of plasma and collision with previously emitted dust cloud, resulting the creation of giant carbon molecule. Nozawa et al.[24)25)] calculated the chemical composition, size distribution, and amount of dust grains formed in the ejecta of population III super novae. After $10^5$ years, there may remain carbon dust with size from 1 to 10 nm, which suggests 100 to 10000 carbons giant molecule. By laboratory experiments[26),27)] and by simulation[28)], it was confirmed that large carbon molecules would be created at a high temperature plasma condition. Those carbon molecules sometimes may react with primally ejected hydrogen cloud, and finally transformed to large size PAHs.

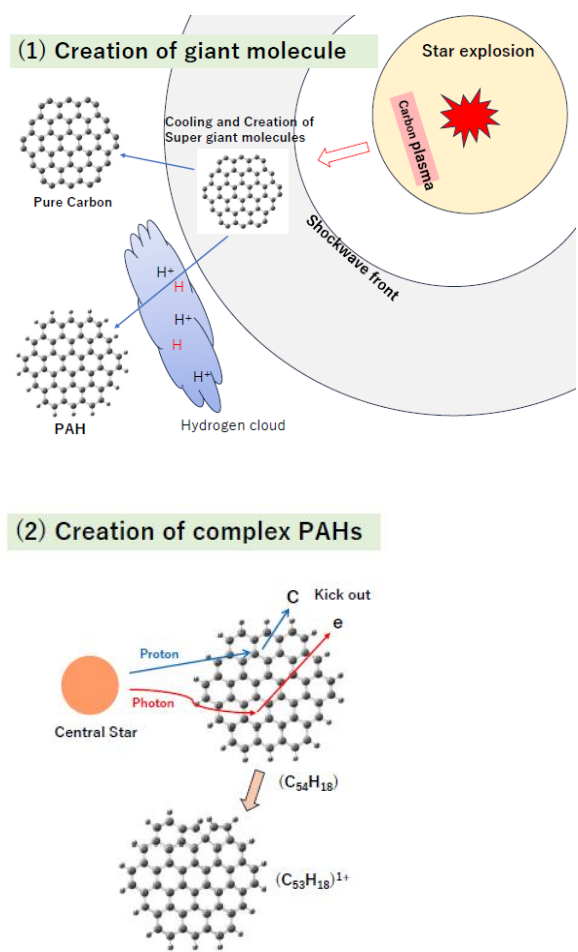

**Fig. 1** (1) Image of creating giant carbon and/or PAH molecules after the star explosion. (2) Creation of pentagon-hexagon combined PAH by high-speed proton attack and high energy photon irradiation from the central star.

A capable second step would be realized around an active central star[29),30)] as shown on bottom in (2). Floating PAH would be attacked by high-speed particles as like proton and/or electrons emitted from central star. Carbon atom in a molecule will be kicked out to result new deformed molecule. Typical example was shown that full hexagon hydrocarbon ($C_{54}H_{18}$) will be transformed to ($C_{53}H_{18}$) having two pentagons among many hexagons. Also, central star illuminates high energy photon on those transformed PAH, which extract electron to result the molecule to the cationic PAH as like mono-cation, di-cation and so on[30)]. Complex PAH and its family would be created by such physical top-down process.

### 3. Calculation Methods

In calculation, we used DFT[31),32)] and TD-DFT[35)] with the unrestricted PBEPBE functional[33), 34)]. We utilized the Gaussian09 software package[35)] employing an atomic orbital 6-31G basis set[36)]. Unrestricted DFT calculation was done to have the spin dependent atomic structure. The required convergence of the root-mean-square density matrix was $10^{-8}$. Based on such optimized molecular configuration, fundamental vibrational modes were calculated. This calculation also gives harmonic vibrational frequency and intensity in infrared region. To each IR spectral line, we assigned a Gaussian profile with a full width at half maximum (FWHM) of 4cm$^{-1}$.

For obtaining the molecular orbital excitation energy and strength, we tried TD-DFT calculation in Gaussian09 package. Excited energies were calculated simultaneously up to 14$^{th}$ excitations.

The scaling factor "s" between the experimental energy to the DFT calculated energy is s=0.990, estimated from fundamental vibration modes of coronene ($C_{24}H_{12}$) discussed by Koichi Ohno in 2011[42)]. Scaling estimation in this study using PBEPBE functional with 6-31G basis set was noted in Appendix-1 and -2.

### 4. Molecule Candidates from IR Study

In our previous papers[2),30)], we tried to find specific PAH molecule under the top-down scheme. We successfully indicated specific molecule group. Typical examples were ($C_{53}H_{18}$) shown in Fig. 2 and ($C_{23}H_{12}$) in Fig. 3. Astronomically observed spectra for four different astronomical objects show general band structure, so called PAH type bands, at 3.2, 6.3, 7.7, 8.6, 11.2, and 12.7µm. We could reproduce those bands by above molecules using DFT calculation.

In large molecule of ($C_{53}H_{18}$) in Fig. 2, charge neutral one show 11.1µm peak close to observed 11.2µm band. Mono-cation reproduced four coincident bands at 3.2, 6.3, 7.6 and 8.7µm. Di-cation calculated bands were somewhat resemble to mono-cation one.

In a small size molecule of ($C_{23}H_{12}$) in Fig. 3, di-cation show good coincidence at 3.2, 7.7, 8.6, 11.2, and 11.6µm. Mono-cation also presents similar behavior.

Those molecules would be candidates reproducing DIB.

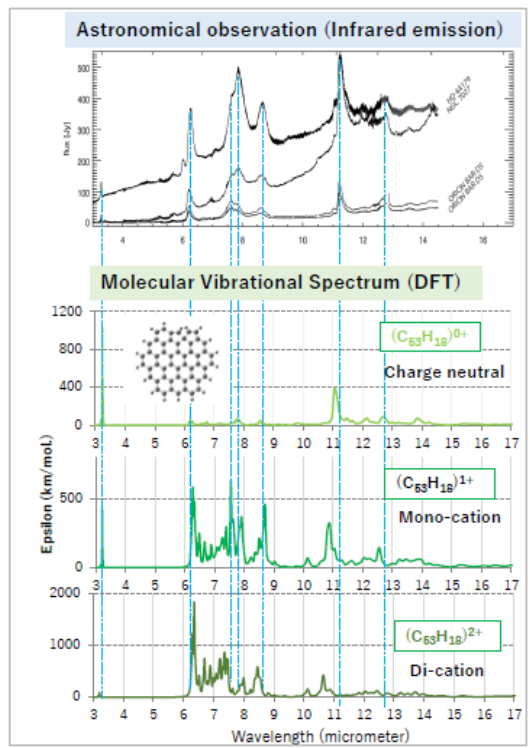

**Fig. 2** Calculated IR of $(C_{53}H_{18})^{n+}$ (charge n=0, 1 and 2) compared with astronomically observed IRs of four different objects.

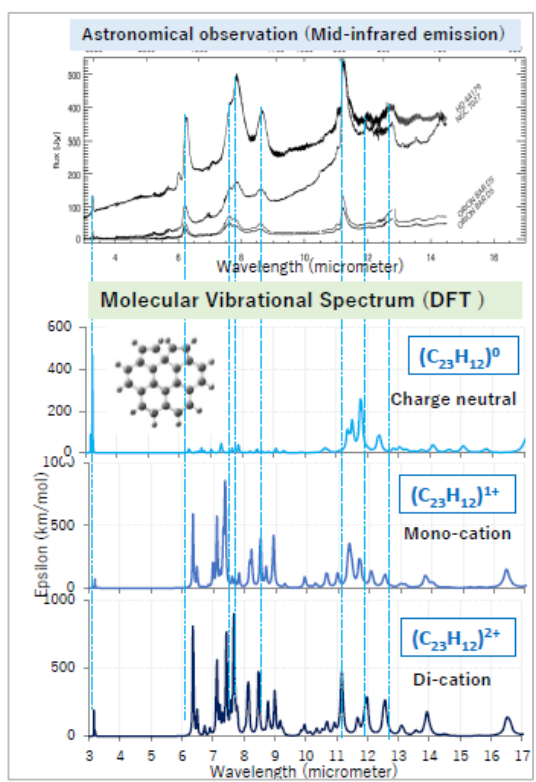

**Fig. 3** Calculated IR of $(C_{23}H_{12})^{n+}$ (n=0, 1 and 2) compared with observed IRs.

## 5. Molecular Orbitals and Excitation Energy

Typical example of molecular orbital calculation was illustrated in Fig. 4 for mono-cation $(C_{53}H_{18})^{1+}$. Optimization of molecular configuration and mapping of spin density was done by DFT. In panel (a), we can see configuration with two hydrocarbon pentagons among hexagon networks. In panel (b), up-spin cloud (by red) and down-spin one (blue) was mapped. Up-spin rich arrangement was appeared around pentagon sites. By TD-DFT calculation, molecular orbital energy diagram was obtained as shown in (c). HOMO orbit was #167 for up-spin alpha bands marked by red arrow and down-spin beta bands by blue. LUMO was #168. We obtained excitation energy among those molecular orbits. Lowest one was absorption is N1 from Beta #167 (HOMO) to #168 (LUMO), which photon energy is N1=2619nm in wavelength. Second one was N2=1567nm from Beta#166 to #168. Third one was N3=1066nm. Other higher excitation energy was calculated until 14th. Molecular orbits were illustrated in (d) for Beta HOMO and LUMO, where orange cloud show plus-sign orbit and green cloud for anti-sign one. LUMO orbit became complex structure than HOMO.

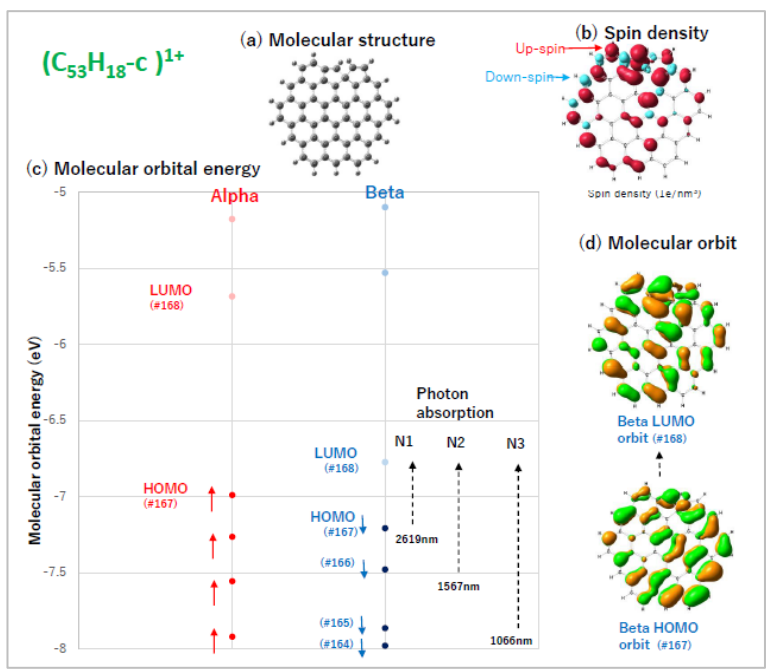

**Fig. 4** Configuration of $(C_{53}H_{18})^{1+}$ in (a), and spin density in (b). TD-DFT calculated molecular orbital energy diagram was shown in (c) for up-spin Alpha bands by red arrows and down spin Beta one by blue. Excitation energy was calculated as N1, N2, N3 up to N14. Molecular orbits were mapped in (d).

## 6. Molecular Orbital Excitation Energy Compared with Observed DIB

Typical example of molecular orbital energy was shown in Fig. 5 for mono-cation $(C_{53}H_{18})^{1+}$. Lan et al.[37] figured observed bands as copied on a top left panel, based on a list by Jenniskens et al.[38]. Our calculation results are listed in Table 1 compared with observed DIB, where FWHM is observed band full width at a half maximum, W is equivalent width related to column density of DIB carriers normalized by E(B-V) of reddening of the astronomical object.

As shown in Table 1, for mono-cation we found one coincide band among 14 calculated bands. Coincide band was marked by yellow. Calculated wavelength matches with observed one within observed FWHM. Calculated excitation number N12 was 627.87nm coincide with observed DIB at 627.83nm, only 0.04nm difference within 0.06nm FWHM.

There reported many longer wavelengths DIB bands as columned in Fig. 5. Suggested mono-cation fullerene $C_{60}^+$ [22],[23], of which bands were reported at 936.5, 942.8, 957.7, 963.2nm. Near and infrared bands were observed by several studies as like Rawlings et al.[39], Geball et al.[40] and Sarre et al.[41]. It should be noted that our example of mono-cation $(C_{53}H_{18})^{1+}$ also show longer wavelengths bands from N1 to N5.

## 7, Large Size Molecule: $(C_{53}H_{18})^{n+}$

In Fig. 6, calculated molecular orbital energy of $(C_{53}H_{18})^{n+}$ were illustrated for every charged state (n=0, 1 and 2) by different green colors. Also, detailed values are listed in Table 1.

In Table 1, charge neutral molecule show 3 coincident bands marked by yellow among 14 bands. Calculated 577.35, 474.24 and 458.11nm correspond to observed 577.95, 476.17 and 459.50nm respectively, within a width of observed FWHM. In observed wavelength column, "n" means not found in the Jenniskens DIB list[38], and "x" means out of the list.

Mono-cation show one coincide bands among 14 calculated bands, of which calculated 627.87nm band correlated to observed 627.83nm. From excitation number N8 to N12, we can see a cluster of bands with difference less than 1nm between calculation and observation.

Di-cation show 3 coincide bands as like calculated 635.89, 615.45 and 505.67nm correspond to 635.95, 617.72 and 503.91nm.

Those many coincide cases were amazing, because until now almost over 100 years, any specified hydrocarbon was not suggested to explain DIB.

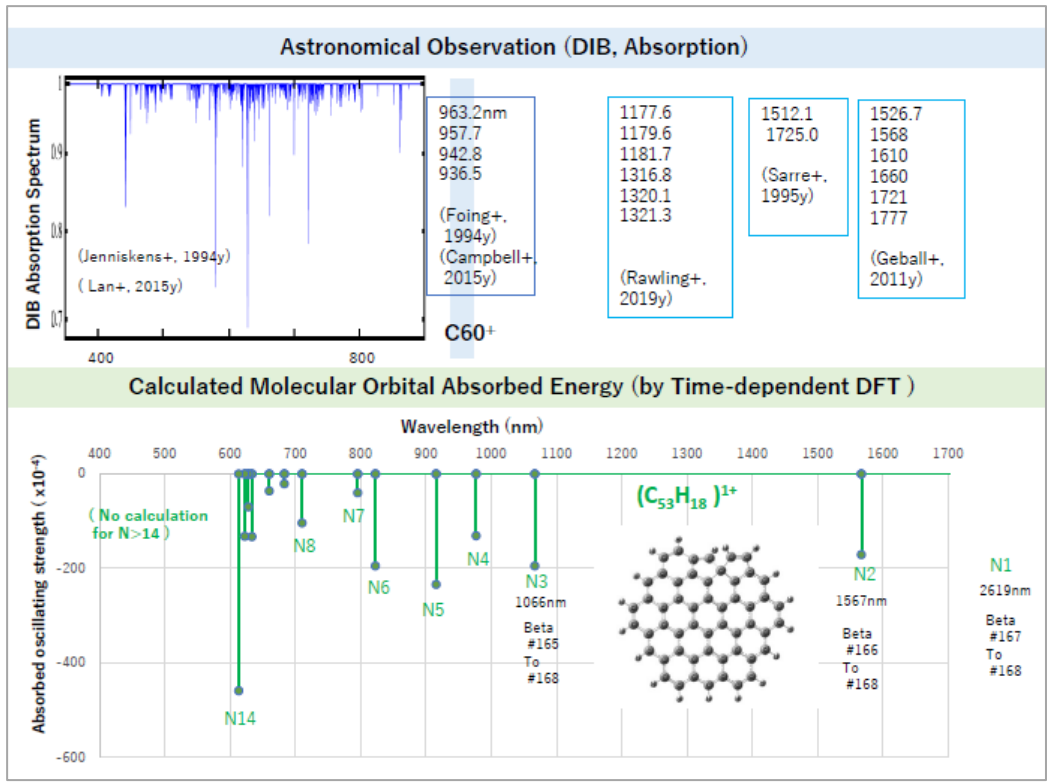

**Fig. 5** Calculated molecular orbital excitation bands for mono-cation $(C_{53}H_{18})^{1+}$ by green lines, compared with astronomically observed DIB bands by blue.

**Table 1,** TD-DFT calculated molecular excitation bands of $(C_{53}H_{18})^{n+}$ (n=0, +1, +2) on left by green columns compared with observed DIB bands [38] by blue. Yellow colored column is coincided band with observed DIB within observed width of FWHM.

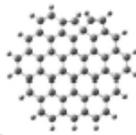

| | | TD-DFT Calculation (This study) S=0.990 | | DIB Observation by Jenniskens et al. [38] | | |
|---|---|---|---|---|---|---|
| | Excitation number | Wavelength (nm) | Oscillating Strength (x10E-4) | Wavelength (nm) | FWHM (nm) | W/EB-V |
| $(C_{53}H_{18})$ Charge neutral | N1 | 761.57 | −485 | 758.56 | 0.09 | 0.01 |
| | N2 | 679.67 | −107 | 679.53 | 0.09 | 0.01 |
| | N3 | 577.35 | −756 | 577.95 | 1.55 | 0.08 |
| | N4 | 558.80 | −960 | 554.50 | 0.08 | 0.02 |
| | N5 | 534.73 | −595 | 536.21 | 0.60 | 0.08 |
| | N6 | 505.57 | −83 | 510.97 | 1.18 | 0.27 |
| | N7 | 491.95 | −601 | 488.18 | 1.97 | 0.61 |
| | N8 | 474.24 | −797 | 476.17 | 2.53 | 0.45 |
| | N9 | 469.43 | −60 | 466.55 | 0.40 | 0.17 |
| | N10 | 458.11 | −1268 | 459.50 | 2.80 | 0.45 |
| | N11 | 452.99 | −2095 | 450.18 | 0.25 | 0.20 |
| | N12 | 441.50 | −165 | 442.89 | 1.23 | 2.23 |
| | N13 | 436.00 | −273 | n (not listed) | | |
| | N14 | 431.17 | −2304 | n | | |
| $(C_{53}H_{18})^{1+}$ Mono-cation | N1 | 2619.26 | −46 | x (out of list) | | |
| | N2 | 1567.07 | −171 | x | | |
| | N3 | 1066.79 | −195 | x | | |
| | N4 | 977.05 | −131 | x | | |
| | N5 | 915.60 | −234 | x | | |
| | N6 | 822.67 | −195 | n | | |
| | N7 | 795.42 | −40 | 792.78 | 1.50 | 0.43 |
| | N8 | 710.27 | −104 | 710.60 | 0.25 | 0.04 |
| | N9 | 683.25 | −21 | 682.72 | 0.10 | 0.02 |
| | N10 | 659.99 | −36 | 659.14 | 0.56 | 0.09 |
| | N11 | 633.54 | −133 | 633.04 | 0.10 | 0.02 |
| | N12 | 627.87 | −70 | 627.83 | 0.06 | 0.02 |
| | N13 | 622.50 | −132 | 622.36 | 0.05 | 0.01 |
| | N14 | 613.19 | −459 | 613.98 | 0.09 | 0.01 |
| $(C_{53}H_{18})^{2+}$ Di-cation | N1 | 2786.96 | −62 | x | | |
| | N2 | 1530.90 | −221 | x | | |
| | N3 | 1012.13 | −594 | x | | |
| | N4 | 916.31 | −279 | x | | |
| | N5 | 851.67 | −218 | 853.08 | 0.17 | 0.07 |
| | N6 | 684.87 | −481 | 684.33 | 0.12 | 0.03 |
| | N7 | 649.71 | −45 | 649.78 | 0.05 | 0.01 |
| | N8 | 635.89 | −631 | 635.95 | 3.73 | 0.54 |
| | N9 | 615.45 | −379 | 617.72 | 2.30 | 0.77 |
| | N10 | 589.42 | −1677 | 585.40 | 0.18 | 0.02 |
| | N11 | 571.41 | −464 | 570.48 | 0.66 | 0.08 |
| | N12 | 523.41 | −311 | n | | |
| | N13 | 505.67 | −1581 | 503.91 | 1.79 | 0.28 |
| | N14 | 500.39 | −298 | 496.97 | 3.37 | 0.50 |

Table 2, Molecular orbital excitation energy of $(C_{23}H_{12})^{n+}$ (left green column) compared with observed DIB bands (right green column).

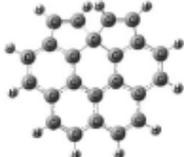

| | | TD-DFT Calculation (This study) s=0.990 | | DIB Observation by Jenniskens et al.[38] | | |
|---|---|---|---|---|---|---|
| | Excitation number | Wavelength (nm) | Oscillating strength (x10E-4) | Wavelength (nm) | FWHM (nm) | W/EB-V |
| $(C_{23}H_{12})^0$ Charge neutral | N1 | 488.16 | −37 | 488.18 | 1.96 | 0.610 |
| | N2 | 449.16 | −13 | 450.18 | 0.25 | 0.200 |
| | N3 | 397.83 | −368 | x | | |
| | N4 | 396.48 | −287 | x | | |
| | N5 | 360.86 | −109 | x | | |
| | N6 | 353.96 | −747 | x | | |
| | N7 | 343.66 | −47 | x | | |
| | N8 | 318.91 | −99 | x | | |
| | N9 | 307.18 | −376 | x | | |
| | N10 | 300.82 | −2695 | x | | |
| | N11 | 290.71 | −21 | x | | |
| | N12 | 290.43 | −296 | x | | |
| | N13 | 290.28 | −154 | x | | |
| | N14 | 280.89 | −1005 | x | | |
| $(C_{23}H_{12})^{1+}$ Mono-cation | N1 | 1932.19 | −77 | x | | |
| | N2 | 1206.18 | −105 | x | | |
| | N3 | 713.92 | −274 | 713.80 | 0.35 | 0.05 |
| | N4 | 608.23 | −43 | 608.98 | 0.08 | 0.02 |
| | N5 | 537.27 | −6 | 536.21 | 0.60 | 0.08 |
| | N6 | 524.39 | −9 | n | | |
| | N7 | 506.84 | −1 | n | | |
| | N8 | 492.60 | −4 | n | | |
| | N9 | 478.18 | −16 | 478.01 | 0.15 | 0.06 |
| | N10 | 476.98 | −97 | 476.17 | 2.53 | 0.62 |
| | N11 | 426.37 | −1 | n | | |
| | N12 | 422.44 | −12 | n | | |
| | N13 | 414.21 | −417 | n | | |
| | N14 | 407.14 | −1 | n | | |
| $(C_{23}H_{12})^{2+}$ Di-cation | N1 | 2144.08 | −62 | x | | |
| | N2 | 1299.37 | −106 | x | | |
| | N3 | 653.27 | −357 | 653.21 | 1.72 | 0.66 |
| | N4 | 552.81 | −106 | 553.70 | 2.26 | 0.29 |
| | N5 | 500.72 | −45 | 503.91 | 1.79 | 0.284 |
| | N6 | 468.98 | −32 | 466.54 | 0.4 | 0.17 |
| | N7 | 435.43 | −549 | n | | |
| | N8 | 428.93 | −73 | n | | |
| | N9 | 419.30 | −54 | n | | |
| | N10 | 402.89 | −20 | n | | |
| | N11 | 382.14 | −32 | x | | |
| | N12 | 362.56 | −1805 | x | | |
| | N13 | 353.38 | −410 | x | | |
| | N14 | 336.84 | −1732 | x | | |

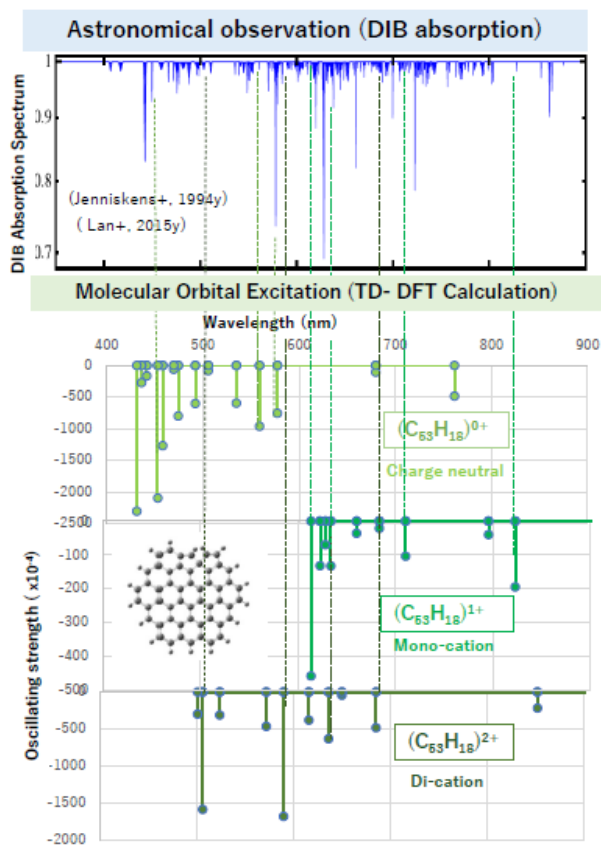

**Fig. 6** Calculated molecular orbital excitation for $(C_{53}H_{18})^{n+}$ (n=0, 1 and 2) by different green lines compared with observed DIB bands by blue lines.

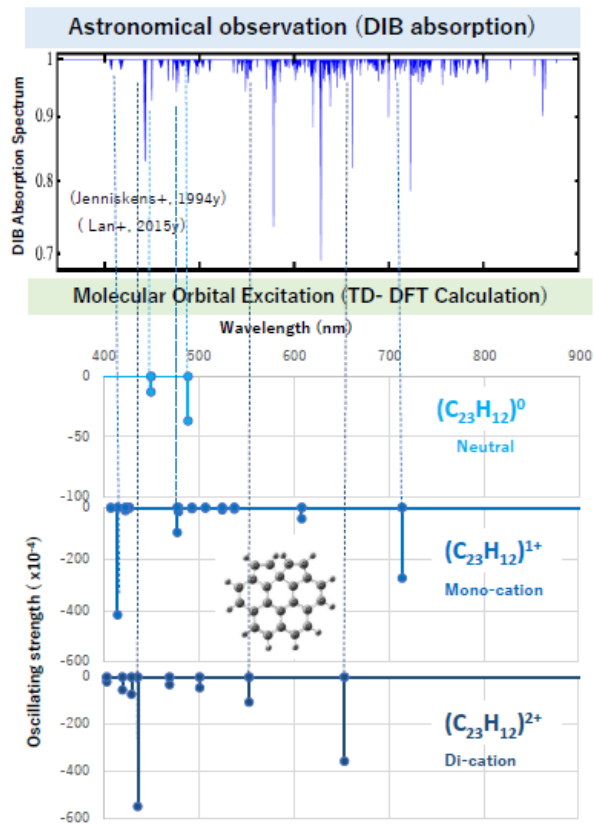

**Fig. 7** Calculated molecular orbital excitation of $(C_{23}H_{12})^{n+}$ (n=0, 1, 2) compared with observed DIB bands.

### 8, Small Size Molecule: $(C_{23}H_{12})^{n+}$

Molecular orbital excitation of small size molecule $(C_{23}H_{12})^{n+}$ (n=0, 1 and 2) are shown in Fig. 7 for every charged state. Detailed comparison with observation[38] was listed in Table 2.

In case of charge neutral, there is one coincident band of N1 at 488.16nm coincide very well with observed 488.18nm with difference only 0.02nm and within width of 1.96nm FWHM. Shorter wavelength excitation less than 400nm were not listed in the Jenniskens observation list[38]. Future observation is necessary in violet wavelength region for small size molecules.

Mono-cation show 2 coincident bands. Calculated N3 at 713.92nm reproduces well observed 713.80nm band within 0.35nm FWHM. Also, N10 at 476.98nm correspond to observed 476.17nm within 2.53nm FWHM.

Di-cation show N3 band at 653.27nm close to observed 653.21nm within 1.72nm FWHM. Also, N4 at 552.81nm was close to observed 553.70nm.

### 9. Conclusion

Purpose of this study is to indicate specific PAH molecules to coincide both astronomically observed IR and DIB bands by calculation.

(1) Under astronomical top-down material creation scheme, we have found the hydrocarbon pentagon-hexagon combined PAH molecules by comparing astronomically observed IR and DFT obtained molecular vibrational spectrum. Molecules were $(C_{53}H_{18})$ and $(C_{23}H_{12})$, which reproducing well, so called PAH type IR bands, at 3.2, 6.3, 7.7, 8.6, 11.2, and 12.7μm.

(2) Origin of DIB may come from the molecular orbital excitation from those molecules. We applied Time-Dependent DFT calculation.

(3) In case of $(C_{53}H_{18})$, we found 7 coincide bands among 42 calculated bands within an accuracy of observed FWHM. For example, neutral $(C_{53}H_{18})$ shows calculated 577.35nm band coincide well with observed DIB at 577.95nm within observed 1.55nm FWHM. Mono-cation of $(C_{53}H_{18})$ show one coincident band at 627.87nm correspond to observed 627.83nm. Di-cation also shows two good coincident bands at 635.89, 615.45 and 505.67nm

related to 635.95, 617.72 and 503.91nm.
(4) For small molecule ($C_{23}H_{12}$), we found 5 coincide bands among 42 calculated bands. Typical example was di-cation molecule's excitation number N3 at 653.27nm coincide well with observed 653.21nm within 1.72nm FWHM.

By such quantum-chemistry survey for reproducing both IR and DIB, we could predict specified PAH molecules floating in interstellar space.

## Acknowledgement

We would like to say great thanks to Prof. Aigen Li of Univ. Missouri for the important suggestions, also thanks to Dr. Christiaan Boersma of NASA Ames Research to give us astronomical IR data. We like to say special thanks to the excellent review paper by Dr. T.R. Geball.

## References

1) T. R. Geballe: J. of Phys.: Conference Series **728**, 062005 (2016)
2) N. Ota, A. Li and L. Nemes: J. Mag. Soc. Japan 45, 86 (2021).
3) A. Tielens: Rev. Mod. Phys. 85, 1021 (2013)
4) J. Oomens: *In PAHs and the Universe*: A Symposium to Celebrate the 25th Anniversary of the PAH Hypothesis, EAS Publications Series (2011)
5) C. Boersma, J. Bregman, and L. Allamandola: *ApJ* **769**, 117 (2013).
6) C. Moutou, K. Sellgren, L. Verstraete, and A. L´eger: A & A, **347**, 949 (1999).
7) E. Peeters, S. Hony, C. van Kerckhoven, et al.: A&A, **390**, 1089 (2002).
8) L.Armus, V.Charmandaris, J.Bernard-Salas, ApJ, **656**,148 (2007).
9) J. Smith, B. Draine, A. Dale, et al.: *ApJ* **656**, 770 (2007).
10) K. Sellgren, K. Uchida and M. Werner: *ApJ* **659**, 1338 (2007).
11) A. Ricca, C. W. Bauschlicher, C. Boersma, A.Tielens & L. J. Allamandola: *ApJ*, 754, 75 (2012).
12) A. Li: *Nature Astronomy*, **4**, 339 (2020).
13) M. L. Heger: Lick Obs. Bull. **10** 146 (1922)
14) G. P. Zwet and L. J. Allamandola: Astron. Astrophys. **146** 76 (1985)
15) A. Leger and L. D'Hendecourt: Astron. Astrophys. **146** 81 (1985)
16) M. K. Crawford, A. Tielens and L. J. Allamandola: Astrophys. J. **293** L45 (1985)
17) H. W. Kroto: Science **242** 1139 (1988)
18) A. L´eger , L. D'Hendecourt , L. Verstraete L and W. Schmidt: Astron. Astrophys. **203** 145 (1988)
19) A. Webster: Mon. Not. R. Astron. Soc. **263** 385 (1933)
20) J. Fulara, M. Jakobi and J. Maier: Chem. Phys. Lett **211** 227 (1933)
21) B. H. Foing and P. Ehrenfreund: Nature **369** 296 (1994)
22) B. H. Foing and P. Ehrenfreund: Astron. Astrophys. **319** L59 (1977)
23) E. K. Campbell, M. Holz, D. Gerlich and J. P. Maier: Nature **523** 322 (2015)
24) T. Nozawa, T. Kozasa, H. Umeda, K. Maeda & K. Nomoto: ApJ, 598,785 (2003)
25) T. Nozawa1 and T. Kozasa: The Astrophysical Journal, **648**, 435 (2006)
26) L. Nemes, A. Keszler, J. Hornkohl, and C. Parigger: *Applied Optics*, **44-18**, 3661 (2005).
27) 27) L. Nemes, E. Brown, S. C. Yang, U. Hommerrich: *Spectrochimica Acta Part A, Molecular and Biomolecular Spectroscopy,* **170**, 145 (2017).
28) N. Patra, P. Kr´al1, and H. R. Sadeghpour: The Astrophysical Journal, **785**, 6 (2014)
29) J. Y. Seok and A. Li: *ApJ,* **835**, 291 (2017).
30) N. Ota nd A. Li: arXiv.org 2303.05645 (2023)
31) P. Hohenberg and W. Kohn: *Phys. Rev.*, **136**, B864 (1964).
32) W. Kohn and L. Sham: *Phys. Rev.,* **140**, A1133(1965).
33) Perdew, J. P.; Burke, K.; Ernzerhof, M. Phys. Rev. Lett. 1996,77, 3865–3868 (1996)
34) Huan Wang, Youjun He, Yongfang Li and Hongmei Su: Phys. Chem. A 2012, 116, 255–262 (2012)
35) M. Frisch, G. Trucks, H. Schlegel et al: Gaussian 09 package software, Gaussian Inc. Wallington CT USA (2009).
36) R. Ditchfield, W. Hehre and J. Pople: *J. Chem. Phys.,* **54**,724(1971).
37) T. Lan, B. Menard and G. Zhu: Mon. Not. R. Astron. Soc. **452** 3629 (2015)
38) P. Jenniskens and F. Desert: 1994 Astron. Astrophys. Suppl. Series **106** 39 (1994)
39) M. G. Rawlings ,1‹ A. J. Adamson ,1,2‹ C. C. M. Marshall 3 and P. J. Sarre: MNRAS **485,** 3398–3401 (2019)
40) T. Geballe, F. Najarro, D. Figer, B. Schlegelmilch, and D. de la Fuente: Nature **479** 200E (2011).
41) P. Sarre, J. Miles, T. Kerr, R. Hibbins, S. Fossey and W. Somerville: Mon. Not. R. Astron. Soc. 277 L41 (1995)

42) K. Ohno: Toyota Research Report **64**, 53 (2011), http://www.toyotariken.jp/activities/64/05ono.pdf/

Authors Profile : Norio Ota, PhD. (太田憲雄)
Magnetic materials and physics, Astrochemistry, Honorary member of the Magnetics Society of Japan, 2010-2021: Senior Professor, Univ. of Tsukuba, Japan. 2003-2011: Exec. Chief Engineer, Hitachi Maxell Ltd.
https://www.researchgate.net/profile/Norio-Ota-Or-Ohta

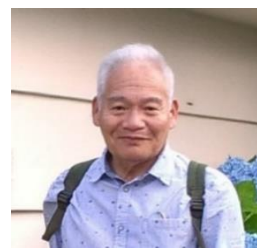

## Appendix-1

Comparison of Molecular vibrational energy for coronene ($C_{24}H_{12}$). Left side blue column show laboratory experimental wavenumbers in cm$^{-1}$ discussed by Koichi. Ohno[42]. Green column table was DFT calculated values under PBEPBE functionals and 6-31G basis set in this study.

| Molecular Vibration Energy (cm$^{-1}$) | | Symmetry | Vibrational mode number |
|---|---|---|---|
| Laboratory Experiment | DFT Calculation PBEPBE 6-31G | | |
| 363.0 | 367.2 | e2g | number-10, 11 |
| 470.6 | 478.5 | a1g | 17 |
| 771.0 | 778.7 | e1u | 36 |
| 992.3 | 998.7 | e2g | 52, 53 |
| 1137.0 | 1149.9 | e1u | 55, 56 |
| 1166.7 | 1178.4 | e2g | 59. 60 |
| 1236.7 | 1253.8 | e2g | 66, 67 |
| 1317.0 | 1328.9 | e1u | 68, 69 |
| 1370.0 | 1393.6 | a1g | 71 |
| 1401.7 | 1418.9 | e2g | 72, 73 |
| 1431.6 | 1461.4 | e2g | 77, 78 |
| 1439.6 | 1465.1 | e2g | 79, 80 |
| 1505.0 | 1515.8 | e1u | 82, 83 |
| 1603.0 | 1612.6 | a1g | 86 |
| 1620.0 | 1628.2 | e1u | 89, 90 |

## Appendix-2

Scaling constant s=0.990 in this study obtained by comparison of molecular vibrational energy noted on Appendix 1.

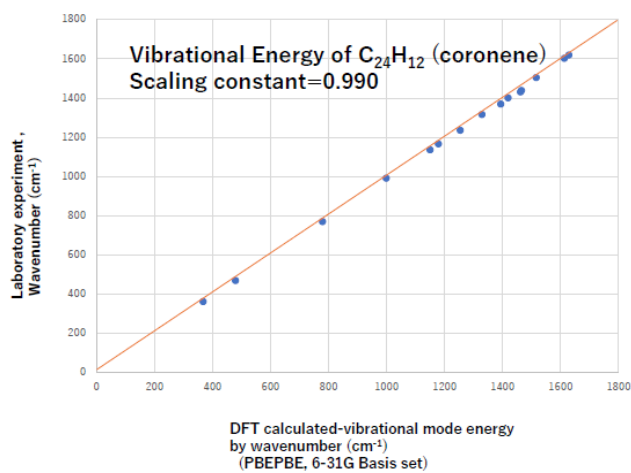